\documentclass[12pt]{article}
\usepackage{graphicx}
\begin{document}
\bibliographystyle{num}
\baselineskip=1.5 \baselineskip
%\draft
\title{Short-range correlations in asymmetric nuclear matter}
\baselineskip=1. \baselineskip

%\author[INP]{Piotr Bo\.zek}
%\address[INP]{Institute of Nuclear Physics, \\
%         Polish Academy of Sciences, PL-31342 Krak\'ow, Poland}
\author{     P. Bo\.{z}ek\footnote{Electronic
address~:
piotr.bozek@ifj.edu.pl}\\
Institute of Nuclear Physics  Polish Academy of Sciences,\\
 PL-31-342 Cracow, Poland}

\date{\today}

\maketitle

\vskip .3cm

%\keywords

\vskip .3cm

\begin{abstract}
The spectral function of protons in the asymmetric nuclear matter is
calculated
in the self-consistent $T$-matrix approach. The spectral function
{\it per proton} increases with increasing asymmetry.
 This effect and the density
dependence of the spectral function partially 
explain the observed
increase of the spectral function with the mass number of the target
nuclei
in electron scattering experiments.
%{ Keywords :}{sum rules, vertex corrections, nuclear matter}
%Nuclear matter, superfluidity, spectral function}
\end{abstract}

%\keywords{ {superfluidity, nuclear matter% saturation point, thermodynamic properties,}
{short-range correlations, electron scattering,  spectral function}
\vskip .4cm

%\pacs
{\bf  21.65.+f, 24.10.Cn, 25.30.-c}

\vskip .4cm

The hard core in the free nucleon-nucleon potential induces strong
short-range correlations in the 
nuclear medium. The inclusion of these
correlations in the modeling of the nuclear medium is crucial 
\cite{brugam}. 
Variational \cite{vcs1,vcs2,fabrocini},   Brueckner-Hartree-Fock 
\cite{EOSBHF,Mahaux,jlm,bhf,Baldoetal92}, and  self-consistent $T$-matrix 
\cite{Dickhoff:1999yi,Bozek:1998su,Dewulf:2003nj,Frick:2003sd} nuclear matter
calculations
take this effect into account. As a result of the
scattering in the medium nucleonic excitations  get dressed.
A measure of such medium modifications
is given by the spectral
function, which can be observed in electron scattering experiments. The
extraction of the spectral function is relatively straightforward in the
Plane Wave Impulse Approximation but important final state
rescattering effects complicate the analysis.
The knowledge of the spectral function  
sheds light on the origin of the nuclear binding energy
\cite{Bozek:2001tz}
 and gives the
nucleon momentum distribution.

 Several calculation of the spectral function exists,
both for finite systems \cite{benharfinite,gfmuther} and for nuclear
matter \cite{fabrocini1,Baldoetal92,fabrocini,Baldo:2001ec,Bozek:2002em}. 
At high momentum and high removal energy the spectral function is
dominated by  short-range correlations which can be studied in
the homogeneous nuclear matter. Existing calculations
are restricted to symmetric or pure neutron nuclear matter. On the
other hand electron scattering experiments are performed on nuclei
with a range of values of the neutron to proton ratio $N/Z$.
 The scaled  spectral
function,
i.e. the spectral function divided
by the number of protons,  was compared for different nuclei and
a significant increase with the mass number was observed \cite{rohe}. 
This is interpreted as an indication of strong short-range correlations due to
neutron-proton scattering which populates the spectral function at
high removal energy \cite{ryckebush}. 
In this work we present the first calculation of the proton spectral
function in the asymmetric nuclear matter. We discuss the effect of
the $N/Z$ ratio  on the
    scaled spectral function for different atomic number of the
    target nuclei.

\begin{figure}
\centering
\includegraphics*[width=0.95\textwidth]{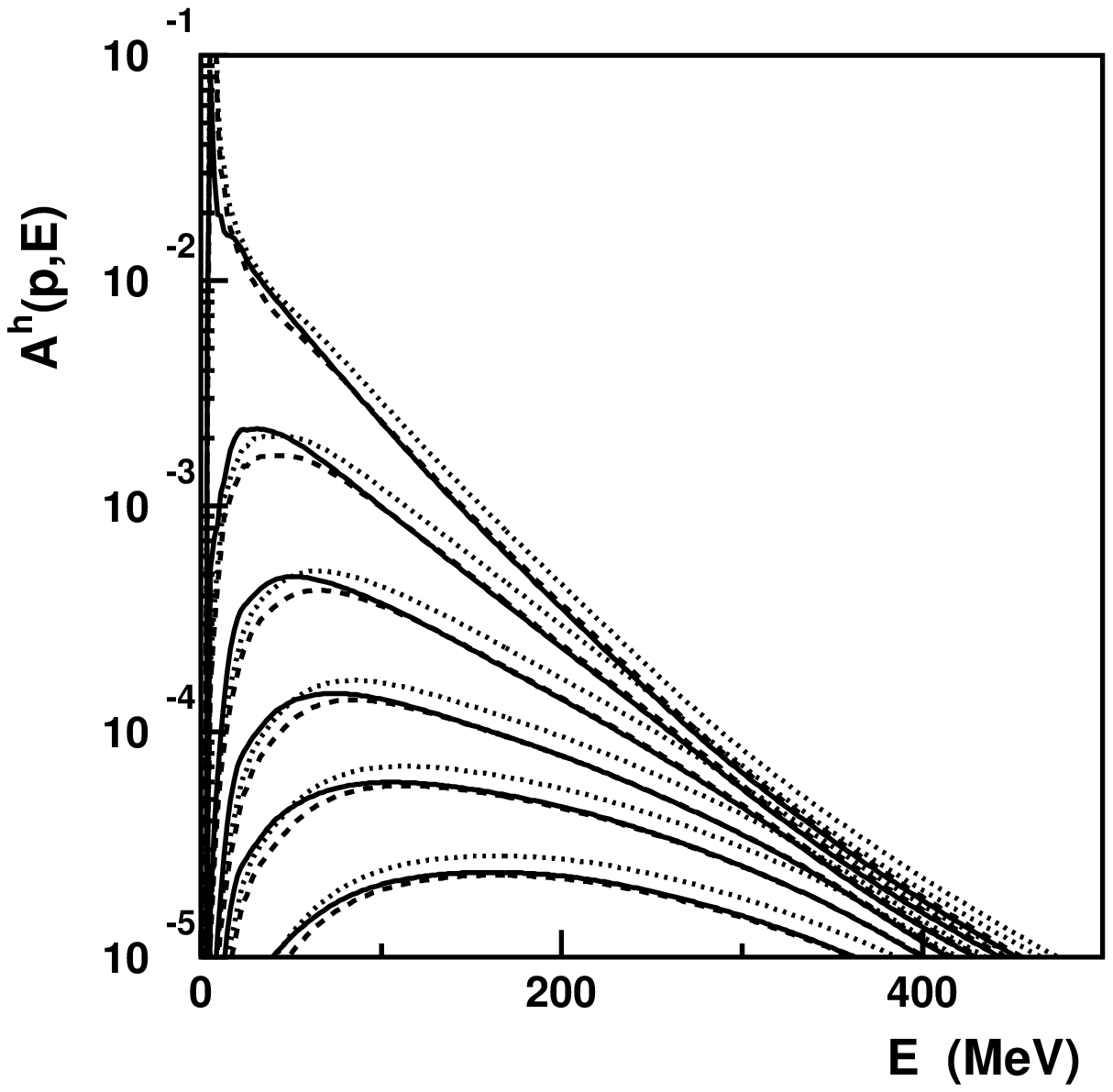}
\caption{The hole spectral function in the  asymmetric nuclear matter at
  normal nuclear density. The lines correspond from top to bottom to 
$p=255, 330, 410, 500, 570, 650$ MeV. The solid lines are for $N=Z$ and
  the dashed lines for $Z/N=0.7$. The dotted lines denote the proton
 spectral function scaled by the mass number
 to proton ratio $\frac{N+Z}{2Z}$, giving a
  measure of the occupancy at a missing energy and momentum {\it per proton}.}
\label{specas}
\end{figure}

The calculation is performed in the self-consistent $T$-matrix
\cite{KadanoffBaym}
approach using fermion propagators dressed in the self-consistent
self-energy. The scheme is thermodynamically consistent and the
obtained single particle properties, i.e. the self-energy and the spectral
function are consistent with the global properties such as the
binding energy \cite{Bozek:2001tz}. In the Brueckner approach the
spectral function can be reliably calculated only after inclusion of
the rearrangement terms \cite{Baldo:2001ec}.
The self-consistent in medium $T$-matrix
\cite{di1,Bozek:1998su,Dickhoff:1999yi,Dewulf:2000jg,Bozek:2001tz} is
\begin{equation}
T=V+ VGGT 
\end{equation}
where the Green's functions $G^{-1}= {\omega-p^2/2m -\Sigma}$ 
in the ladder propagator
are dressed
by the self-consistent self-energy
\begin{equation}
i\Sigma= Tr[TG]  \ .
\end{equation}
The details of the calculation can be found in Refs.
\cite{Bozek:2002em,Bozek:2002tz}.
The same retarded self-energy $\Sigma$ is used to construct the
spectral function
\begin{eqnarray}
A(p,\omega)&=&-2 {\rm Im} G(p,\omega) \nonumber \\
&=&\frac{-2{\rm Im}\Sigma(p,\omega)}{
(\omega-p^2/2 m-{\rm Re} \Sigma(p,\omega)+\mu)^2+{\rm Im}
\Sigma(p,\omega)^2 } \ .
\label{normalspec}
\end{eqnarray}
The calculation is done for a range of densities
and asymmetries
% between $0.2$ and $1$
%normal nuclear density $\rho_0=0.16$fm$^{-3}$ 
using the CD-Bonn
potential. 
%The neutron-proton
%asymmetry ranges from $Z/N=1$ to $0.7$. 
The hole spectral function $A^h(p,E)=A(p,-E)$ is shown in Fig.
\ref{specas} as a function of the missing energy for several values of
the momentum. The spectral function  at normal nuclear
density is presented for a symmetric nuclear matter and for
$Z/N=0.7$. The proton spectral function
is larger for the symmetric nuclear matter,
% which is not surprising
%since the proton density is a factor $1.25$ larger. 
but at large
missing energy the two curves almost coincide.
 Since in  the asymmetric medium a lower number of p-p collisions is
compensated by a more frequent  n-p collisions  contribution, 
tensor interaction would
lead to a higher scattering rate for high momentum on-shell protons in
the asymmetric medium. A far off-shell proton however is
scattered  in a similar manner in the symmetric and
asymmetric nuclear matter. In fact $|Im \Sigma(p,\omega)|$ for protons
far off-shell ($\omega<\mu$, $p>p_F$)
is even slightly smaller in the asymmetric  than in
the symmetric nuclear matter.  Far off-shell the
value of the 
spectral function is determined by the imaginary part of the self-energy in
the numerator of Eq. \ref{normalspec}, and leads to almost no
variation of the hole spectral function with asymmetry.
Only for energies closer to the quasi-particle
pole ($p>p_F$,  $\omega>\mu$) is the scattering of protons larger in the
asymmetric nuclear matter, giving a higher on shell width for high
momentum protons in the neutron rich nuclear matter. 

When comparing different proton densities in the asymmetric nuclear matter
 the scaled
spectral function $\frac{N+Z}{2Z} A^h(p,\omega)$ is used, which is a
 measure of the spectral function per proton.
 We see that the high energy
and
high momentum part of the scaled hole spectral function is larger in the
asymmetric nuclear matter. The increase of the far off-shell
 correlations {    per proton} compensates the slight decrease of the proton
off-shell scattering rate.

\begin{figure}
\centering
\includegraphics*[width=0.5\textwidth]{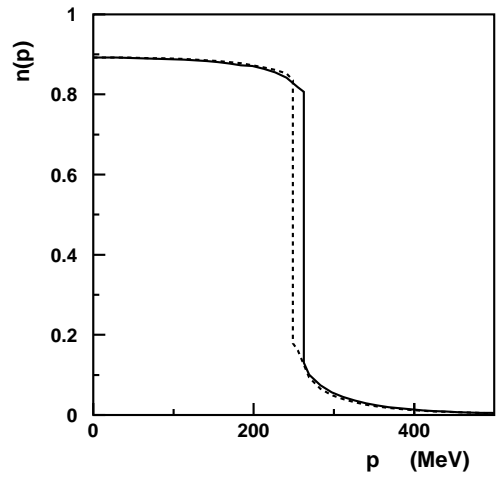}
\caption{The momentum distribution of protons in the nuclear matter at
  normal density. The solid line is for $N=Z$ and
  the dashed line for $Z/N=0.7$ }
\label{nk}
\end{figure}

The real-part of the self-energy and the quasiparticle poles
 depend on the neutron to proton ratio.
Accordingly the proton Fermi momentum depends on the proton density,
which changes with the proton content. In Fig. \ref{nk} is plotted the
proton momentum distribution
\begin{equation}
n(p)=\int \frac{d E}{2\pi} A^h(p,E) \ .
\end{equation}
 in nuclear matter at normal density for
different $Z/N$ ratios. The depletion of the  occupation number inside the
Fermi sphere and the high momentum component in the distribution are
given by  short-range correlations in the medium. These effects to
within a few percent are the same for the  proton to neutron ratio changing
from $1$ to $0.7$. The change in the proton density with asymmetry 
 is accommodated by a
shift in the proton Fermi momentum.

\begin{figure}
\centering
\includegraphics*[width=0.95\textwidth]{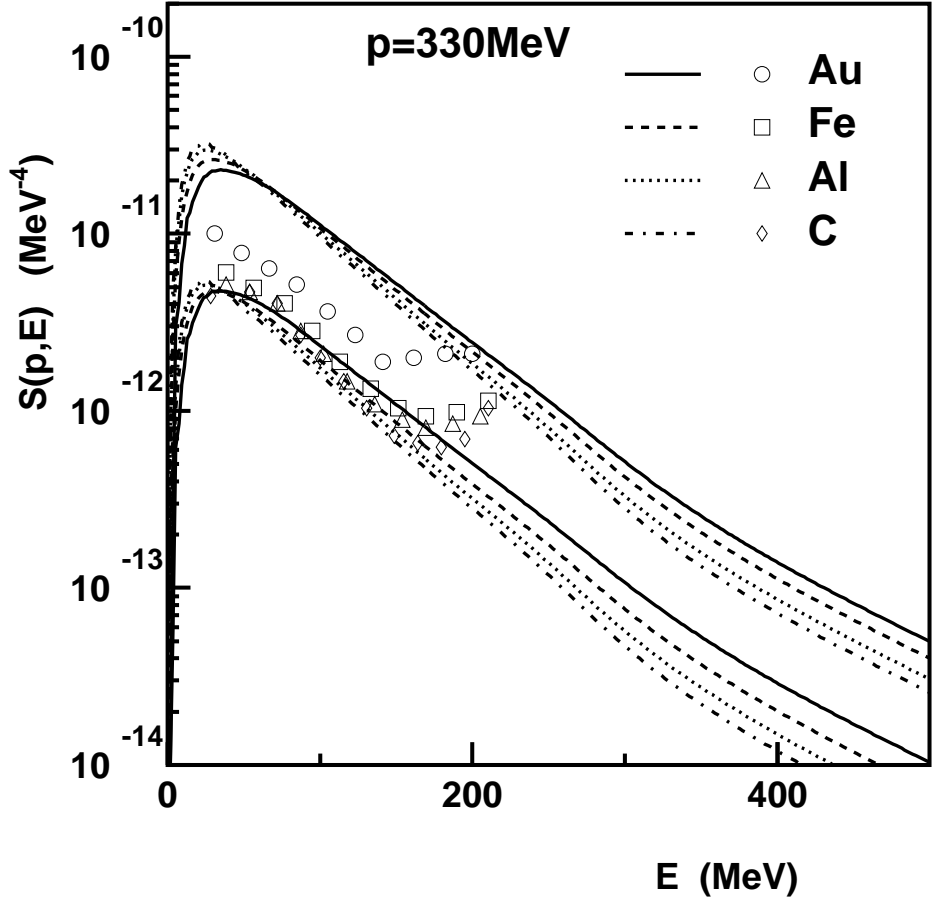}
\caption{The hole spectral function scaled by the number of protons for
Au, Fe, Al and C nuclei; solid, dashed, dotted, and dashed-dotted lines,
respectively (lower curves). The upper curves are multiplied by $\frac{12
  Z}{A}$. The points represent the data from Ref. \cite{rohe} in the
region below the $\Delta$ resonance.}

\label{sp}
\end{figure}

Experiments with electron scattering on finite nuclei are available
for several
target masses (C, Al, Fe and Au) \cite{rohe}. A strong dependence on the
target of the scaled proton spectral function
 was interpreted partly as a signal
of strong p-n correlations. The increase of the off-shell hole
spectral function per proton when going from C to Au target is about a
factor $2$.
% in the region outside of the quasiparticle peak  or  the
% region where $\Delta$ contributes.
The spectral function at high missing energy and momentum can be
calculated using
the local density approximation
\begin{equation}
S(p,E)=\frac{2}{(2\pi)^4}\int d^3 r A^h(p,E,\rho(r)) \ .
\end{equation}
In Fig. \ref{sp} is plotted the spectral function divided by the
number of protons for several nuclei. The spectral function depends on
the mass number of the target. Part of the increase of the spectral
function at high missing energy is due to the different proton to
neutron ratio. We have
\begin{equation}
S(p,E) \propto {\rm Volume} \ A_h(p,E) =\frac{Z}{\bar
  {\rho}_p}
A_h(p,E) 
\end{equation}
where $\bar{\rho}_p$ is the mean proton density.
The scaled spectral function is
 \begin{equation}
\label{ssca}
\frac{S(p,E)}{Z}  \propto \frac{1}{\bar
  {\rho}_p}A_h(p,E) = \frac{1}{\bar{\rho}} \frac{N+Z}{Z} A_h(p,E) \ .
\end{equation}
Since the hole spectral functions  $A_h(p,E)$ at different proton 
densities  are 
 very similar in the considered region of energy and momentum
 (Fig. \ref{specas}) we see from Eq. \ref{ssca} that the scaled proton
 spectral function $S(p,E)/Z$ increases with the asymmetry for nuclei with
 the same average density $\bar{\rho}$.  
This effect however is not sufficient to explain quantitatively the
observed dependence on the 
 target mass.
 
The upper curves in 
 Fig. \ref{sp} are the spectral functions without dividing by
 $Z$, i.e. the lower curves multiplied by $12 Z/A$. In a homogeneous
 matter this rescaling should bring the curves back onto each other as
 in Fig \ref{specas}. The remaining difference visible in
 Fig. \ref{sp}
 is due to the density
 dependence of the spectral function. Heavier nuclei have a larger
 mean density which gives rise to stronger short-range correlations.
%In fact the proton to neutron ratio is almost the same in Al and C nuclei.
Qualitatively the same effect is seen in the data \cite{rohe}, but
the strongest increase of the proton spectral function happens between
Fe and Au targets, and the effect is much more pronounced, especially
in the region where the $\Delta$ resonance starts to be visible.

The proton spectral function in the asymmetric nuclear matter is
calculated. In the far off-shell region the spectral function does
not depend strongly on the proton-neutron asymmetry. It means that the 
spectral function scaled by the proton number is approximately
inversely proportional to the proton density (\ref{ssca}). 
This effect together with
density dependence of the spectral function leads to an increase of
the scaled proton spectral function with increasing mass number of
the target. The experimentally observed increase of the spectral
function for heavy target nuclei is stronger than our estimate, especially for
the case of  gold.  Larger
frequency of  neutron-proton interactions in asymmetric matter
leads to a stronger
scattering of proton excitation on-shell, but 
in the off-shell region ($p>p_F$, $\omega<\mu$) we
find that the scattering width $2|Im\Sigma(p,\omega)|$ decreases
slightly with the asymmetry.

{\bf Acknowledgments}
\vskip .3cm
This work was partly supported by the KBN
under Grant No. 2P03B05925. The author thanks the organizers of the
ECT$^\star$ Workshop on ``Contributions of Short and Long-range
correlations
to Nuclear Binding and Saturation'' in Trento for a fruitful
atmosphere and Daniela Rohe for an illuminating discussion. 

\bibliography{../mojbib}

%\newpage

\end{document}